\documentclass[letterpaper, 10 pt, conference]{ieeeconf} 
\usepackage{graphicx}
\IEEEoverridecommandlockouts         
\usepackage{xcolor}
\usepackage{comment}
\usepackage{float}
\usepackage{lipsum}
\let\labelindent\relax
\usepackage{enumitem}
\setlist[enumerate]{leftmargin=*}
\newtheorem{assumption}{Assumption}
\newtheorem{definition}{Definition}
\newtheorem{lemma}{Lemma}

\newtheorem{theorem}{Theorem}

\newtheorem{remark}{Remark}
\usepackage{comment}
\usepackage{amsmath}
\usepackage{amssymb}
\usepackage{mathtools}

\DeclareMathOperator{\diag}{diag}

\newcommand{\at}{\tilde{A}}
\newcommand{\bt}{\tilde{B}}

\def\D{\mathcal{D}}
\def\X{\mathcal{X}}
\def\T{\mathcal{T}}
\def\R{\mathcal{R}}
\def\C{\mathcal{C}}
\newcommand{\OO}{\mathcal{O}}
\newcommand{\RR}{\mathbb{R}}
\newcommand{\NN}{\mathbb{N}}
\newcommand{\p}{\circ}
\usepackage{bm}

\title{\LARGE \bf
A Unified KKL-based Interval Observer for Nonlinear Discrete-time Systems}

\author{Thach Ngoc Dinh$^*$ and Gia Quoc Bao Tran$^*$ 
\thanks{$^*$The authors contributed equally to this work. Their names are listed in alphabetical order.}
\thanks{Thach Ngoc Dinh is with Conservatoire National des Arts et Métiers (CNAM), Cedric-Lab, 292 rue St-Martin, 75141 Paris Cedex 03, France (e-mail: ngoc-thach.dinh@lecnam.net). Gia Quoc Bao Tran is with Centre Automatique et Systèmes (CAS), Mines Paris - PSL, Paris, France (e-mail: gia-quoc-bao.tran@minesparis.psl.eu).}
}

\begin{document}

\maketitle
\thispagestyle{empty}
\pagestyle{empty}


\begin{abstract}
This work proposes an interval observer design for nonlinear discrete-time systems based on the Kazantzis-Kravaris/Luenberger (KKL) paradigm. Our design extends to generic nonlinear systems without any assumption on the structure of its dynamics and output maps. Relying on a transformation putting the system into a target form where an interval observer can be directly designed, we then propose a method to reconstruct the bounds in the original coordinates using the bounds in the target coordinates, thanks to the Lipschitz injectivity of this transformation achieved under Lipschitz distinguishability when the target dynamics have a high enough dimension and are pushed sufficiently fast. An academic example serves to illustrate our methods.
\end{abstract}

\section{Introduction}
\label{sec:introduction}
The concept of interval observers traces back to the pioneering work of Gouzé et al. in 2000 \cite{Gouze-00}. Since then, it has evolved in various directions, driven by the crucial role of state estimation in monitoring, fault detection, and control applications (for more detailed explanations, refer to \cite{Efimov-16} and the cited references). In essence, interval observers bound the actual state between two functions at each time instant. While this design approach has proven successful, it does come with the cost of certain assumptions. Indeed, a key feature of interval observers is that they can be constructed when the initial conditions as well as the uncertainties are upper and lower bounded by known vectors, and the interval property requires a direct or indirect notion of a non-negative and cooperative system.  

In cases involving nonlinear dynamics, interval observers have been proposed in various works. It is crucial to highlight that to the best of our knowledge, it appears that all existing works focus on nonlinear systems with assumptions about the functions of the state and/or output. For instance, some papers such as \cite{8431441, Dinhetal:2023:IFACWC} assume a specific structure of the dynamics map, in particular, a linear part providing observability followed by a Lipschitz nonlinearity. On the other hand, the work in \cite{Khajenejadetal:2023:CDC} supposes that the state maps and output maps have bounded Jacobians with known/computable Jacobian bounds, alongside the existence of a Jacobian sign-stable decomposition of the output maps. In \cite{mohamad-20}, the requirement is that the vector fields of the state and output are mixed monotone. In \cite{Efimov-13}, it is necessary for the values of the nonlinear state functions to be enclosed within a known interval. Lastly, \cite{thachecc} assumes that the family of nonlinear systems is affine in the unmeasured part of the state variables. 

On the other hand, the Kazantzis-Kravaris/Luenberger (KKL) observer is a powerful universal theory for observer design. This method consists of transforming the given nonlinear dynamics into some target dynamics, where a simple observer is designed giving us exponential stability of the estimation error in the new coordinates, and inverting this transformation to recover the estimate. The property (equivalent or weaker than exponential stability) of the error that we can bring back to the original coordinates depends on the injectivity of the transformations, which in turn relies on the observability of the original system. This design has been proposed for various classes of systems, both in continuous \cite{andrieuPralyKKL,kklParameter,brivadisRemarks} and discrete time \cite{brivadisCDC,baokkltac} and the references therein. The advantage of this design is its genericity, being applicable to (possibly time-varying) nonlinear systems of structure-free dynamics and output maps by gathering all the nonlinearity and time variation into the transformation. Consequently, the closed forms of these transformations become very difficult to compute in practice, leading to the development of AI tools to learn those maps and their inverses from data \cite{BuiBahDiM}, thus a unified framework for asymptotic observer design for essentially any systems.

In this paper, we develop a unified interval observer design framework for nonlinear discrete-time systems, based on the KKL spirit. Exploiting the robustness of the KKL observer in \cite{baokkltac} and the properties of the KKL transformations, we can construct an interval observer (instead of an asymptotic one) in the target coordinates and then reconstruct the bounds in the original coordinates. It is worth noting that the effectiveness of traditional nonlinear observers (such as KKL, LMI,...) in monitoring and detecting system faults during transient periods is limited. This limitation highlights the advantage of interval observers over classical methods, particularly in providing guarantees for monitoring and detection in dynamic environments. To the best of our knowledge, while satisfactory solutions exist in specific cases, interval observers for nonlinear systems still lack generality, and there is no unified and systematic method for the design of such filters. The objective of the present work is to encompass a broader class of nonlinearities compared to existing approaches and to address the challenge of designing an interval observer for nonlinear systems without any prior knowledge of the structure of the system's dynamics and output maps. The preprint \cite{KKLHugoetal} also aims for a similar objective, albeit with the assumption that the inverse transformation is differentiable to ensure mixed monotonicity, along with an additional requirement of a non-negative target dynamics matrix. It is noteworthy that our main results do not hinge on the inverse transformation being mixed monotone, and the choice of the target dynamics matrix does not necessitate non-negativity. Therefore, this work serves as a first milestone for more in-depth follow-up research in this direction in the future. 
 
\textit{Notations:} We use standard notations, which are simplified when no confusion arises from the context. The inequalities such as $a \leq b$ for vectors $a$, $b$ or $A \leq B$ for matrices $A$, $B$ are component-wise. For a matrix $M \in \RR^{n \times m}$ with entries $m_{i,j}$, define $M^\oplus$ as the matrix in $\RR^{n \times m}$ whose entries are $\max\left\{0, m_{i, j}\right\}$ and let $M^\ominus = M^\oplus - M$. For a scalar $x \in \RR$, the absolute value of $x$ is denoted by $|x|$. For a vector $x\in \RR^{n_x}$, its $\infty$-norm\footnote{For illustration, the $\infty$-norm is used in this paper. However, our results hold for any norm thanks to their equivalence (in a finite-dimensional space).} is $\|x\|:=\displaystyle\max_{i=\overline{1,n_x}}\left|x_i\right|$, where $x_i$ is the $i^{\rm th}$ component of $x$. Similarly, $x_{k,i}$ is the $i^{\rm th}$ component of $x_k$. Denote $E_n \in \RR^{n}$ as the vector whose entries are all $1$.

In this work, we design interval observers as defined next.
\begin{definition}\label{def_intobs}
Consider the nonlinear discrete-time system
\begin{equation}\label{eq:defint}
       x_{k+1} = f\left(x_k\right) + d_k, \qquad
        y_k = h\left(x_k\right) + w_k,
\end{equation}
with $x_k \in \mathbb{R}^{n_x}$, $y_k \in \mathbb{R}^{n_y}$, $d_k \in \mathbb{R}^{n_x}$, $w_k \in \mathbb{R}^{n_y}$, and where $f$ and $h$ are two functions. The uncertainties $(d_k)_{k \in \NN}$ and $(w_k)_{k \in \NN}$
are such that there exist known sequences $(d_k^+,d_k^-,w_k^+,w_k^-)_{k \in \NN}$
such that for all $k \in \NN$, $d_k^- \leq d_k \leq d_k^+$ and $w_k^- \leq w_k \leq w_k^+$.
Moreover, the initial condition $x_0 \in \mathbb{R}^{n_x}$ is assumed to be
bounded by two known bounds: $x_0^- \leq x_0 \leq x_0^+$.
Given a transformation $x_k \mapsto z_k = T(x_k)$ with $T: \X \subset \RR^{n_x} \to \RR^{n_z}$ and $T^*: \RR^{n_z} \to \RR^{n_x}$ an inverse map of $T$, the following dynamics for all $k \in \NN$
\begin{subequations}
\begin{align}
\hat{z}^+_{k+1}&=\mathcal{\overline Z}(k,\hat{z}^+_k,y_k,d^+_k, d^-_k, w^+_k,w^-_k),
\\
\hat{z}^-_{k+1}&=\mathcal{\underline Z}(k,\hat{z}^-_k,y_k,d^+_k, d^-_k, w^+_k,w^-_k),
\end{align} 
associated with the initial conditions
\begin{align}
\hat{z}^+_0&=\mathcal{\overline{Z}}_0(T(x^+_0),T(x^-_0), x^+_0,x^-_0), \\\hat{z}^-_0&=\mathcal{\underline{Z}}_0(T(x^+_0),T(x^-_0), x^+_0,x^-_0),
\end{align}
and the outputs for all $k\geq 1$
\begin{align}\label{eq:io}
x_k^+&=\mathcal{\overline X}(k, T^*(\hat{z}_k^+),T^*(\hat{z}^-_k), \hat{z}_k^+,\hat{z}^-_k),
\\
x_k^-&=\mathcal{\underline X}(k, T^*(\hat{z}_k^+),T^*(\hat{z}^-_k), \hat{z}_k^+,\hat{z}^-_k),
\end{align}
\end{subequations}
for some maps $(\underline{\mathcal{Z}},\overline{\mathcal{Z}},\underline{\mathcal{Z}}_0,\overline{\mathcal{Z}}_0,\underline{\mathcal{X}},\overline{\mathcal{X}})$, are called a \textit{KKL-based interval observer} for system \eqref{eq:defint}
if:
\begin{enumerate}
\item $x_k^- \leq x_k \leq x_k^+$ for all $k\geq 1$;
\item $\displaystyle \lim_{k\to+\infty} \|x_k^+ - x_k^-\| = 0$ when $d_k=0$ and $w_k=0$ $\forall k\in\NN$.
\end{enumerate} 
\end{definition}

Note that below we remove the disturbance $d_k$ from the dynamics for simplicity, without losing the generality in Definition \ref{def_intobs} (see Remark \ref{rem}). The following mathematical results are needed for understanding this paper.
\begin{lemma}\cite[Section II.A]{Efimov-12}
\label{lem_pm}
Consider vectors $a$, $a^+$, $a^-$ in $\mathbb{R}^{n}$ such that $a^- \leq a \leq a^+$. For any $A \in \RR^{m\times n}$,
\begin{equation}\label{eq:lem2}
A^{\oplus}a^- -A^{\ominus}a^+ \leq Aa \leq A^{\oplus}a^+ -A^{\ominus}a^-.
\end{equation}
\end{lemma}
\begin{lemma}\label{lem_Rk}\cite[Theorem 4]{mazenc2014interval}
    For any $A$ Schur, there exist a sequence of invertible real matrices $(R_k)_{k \in \NN}$ and some $\sigma>0$ such that for all $k \in \NN$, $\left\|R_k\right\|+\left\|R_k^{-1}\right\| \leq \sigma$ and $R_{k+1}AR_k^{-1}$ is a non-negative Schur \textit{constant} matrix. 
\end{lemma}

\section{Main Results}\label{sec_nonlin}
\subsection{Problem Statement}
Consider a nonlinear discrete-time system
\begin{equation}\label{eq:sysx}
       x_{k+1} = f(x_k), \qquad
        y_k = h(x_k) + w_k,
\end{equation}
where $x_k \in \RR^{n_x}$ is the state, $y_k \in \RR^{n_y}$ is the measured output, and $(w_k)_{k \in \NN}$ is the sequence of measurement noise. Some assumptions are then made for system \eqref{eq:sysx} as follows. 
\begin{assumption}\label{ass_sys}
    For system \eqref{eq:sysx}, we assume that:
\begin{enumerate}[label=(A\arabic*)]
        \item \label{ass_x} There exist sets $\X_0 \subset
        \X \subset \RR^{n_x}$ such that for all $x_0 \in \X_0$, $x_k \in \X$ for all $k \in \NN$, where $\X_0$ is a subset of $\left[x_0^-,x_0^+\right]$ with $x_0^-$ and $x_0^+$ known;
        \item \label{ass_inv} The map $f$ is invertible as $f^{-1}$ that is defined everywhere;
        \item \label{ass_fh} There exist $c_f > 0$ and $c_h > 0$ such that for all $(x_a,x_b) \in \RR^{n_x} \times \RR^{n_x}$, we have
        \begin{subequations}
            \begin{align}
            \|f^{-1}(x_a) - f^{-1}(x_b)\| & \leq c_f\|x_a - x_b\|,\\
            \|h(x_a) - h(x_b)\|& \leq c_h\|x_a - x_b\|;
        \end{align}
        \end{subequations}
        \item \label{ass_obs} System \eqref{eq:sysx} is Lipschitz backward distinguishable on $\X$ for some positive $m_i \in \NN$, $i \in \{1, 2, \ldots, n_y\}$, and $c_o > 0$ (see below in Definition \ref{def_obs}).
    \end{enumerate}
\end{assumption}

\begin{definition}\label{def_obs}
System \eqref{eq:sysx} is \textit{Lipschitz backward distinguishable} on a set $\X$  if for each $i \in \left\{1, 2, \ldots, n_y\right\}$, there exists a positive $m_i \in \NN$ such that the backward distinguishability map $\OO$ defined as
\begin{subequations}
    \begin{equation}
        \OO(x) = \left(\OO_1(x), \OO_2(x), \ldots, \OO_{n_y}(x)\right),
    \end{equation}
where $\OO_i(x) \in \RR^{m_i}$ is defined as
\begin{equation}
    \OO_i(x) = \begin{pmatrix*}[l]
        (h_i \p f^{-1})(x) \\
        (h_i \p f^{-1} \p f^{-1})(x) \\
        \ldots \\
        (h_i \p \overbrace{f^{-1} \p \ldots \p f^{-1}}^{m_i \text{ times}})(x)
    \end{pmatrix*},
\end{equation}
\end{subequations}
is Lipschitz injective on $\X$, i.e., there exists $c_o > 0$ such that for all $(x_a,x_b)\in \X \times \X$,
\begin{equation}
    \|\OO(x_a) - \OO(x_b)\| \geq c_o\|x_a - x_b\|.
\end{equation}
\end{definition}

\begin{remark}
    While the properties in Items \ref{ass_inv} and \ref{ass_fh} of Assumption \ref{ass_sys} are for now required globally, since the true solution $x_k$ is known to remain in the set $\X$, it is possible to modify the observer outside of a slightly bigger bounded set containing $\X$. Then, the constants in Item \ref{ass_fh} of Assumption \ref{ass_sys} are taken on this slightly bigger set instead of $\RR^{n_x}$, thus reducing conservativeness (see \cite[Section IV.D]{baokkltac} for more details). Note also that Definition \ref{def_obs} gives a property that is stronger than the \textit{backward distinguishability} in \cite{brivadisCDC} by a Lipschitz constant, because we later rely on this for the Lipschitz injectivity of the KKL transformation, resulting in exponential stability (rather than asymptotic stability) of the error and allowing us to reconstruct the interval observer bounds in the original coordinates.
\end{remark}

The objective here is to build for system \eqref{eq:sysx} an interval observer as is Definition \ref{eq:defint}.
Following the KKL paradigm \cite{brivadisCDC}, we strive for a transformation $x_k \mapsto z_k = T(x_k)$ with $T: \X \to \RR^{n_z}$ satisfying
\begin{equation}\label{eq:sylvester}
    T(f(x)) = AT(x) + Bh(x), \qquad \forall x \in \X.
\end{equation}
Thanks to Item \ref{ass_x} of Assumption \ref{ass_sys}, where $A$ is Schur and $(A,B)$ is controllable, through which system \eqref{eq:sysx} is put into
\begin{align}\label{eq:sysz}
        z_{k+1} = A z_k + B y_k - Bw_k.
\end{align}
Our method revolves around the design of an interval observer in the coordinates of system \eqref{eq:sysz}. This observer provides us with bounds on $z_k$. Subsequently, by leveraging the Lipschitz injectivity of $T$ and the robustness inherent in the KKL design, we derive the corresponding bounds for $x_k$.

\subsection{Properties of $T$}
In this part, we summarize the properties of the map $T$ that are useful later for observer design.
\begin{lemma}\label{lemma}
Suppose Assumption \ref{ass_sys} holds. Define $n_z = \sum_{i = 1}^{n_y} m_i$ and denote $\overline{m} = \max_{i=\overline{1,n_y}} m_i$. Consider for each $i$ in $\{1, 2, \ldots, n_y\}$ a controllable pair $(\at_i, \bt_i) \in \mathbb{R}^{m_i\times m_i} \times \mathbb{R}^{m_i}$ where $\at_i$ is Schur. There exists $\gamma^\star \in (0,1]$ such that for any $0 < \gamma < \gamma^\star$, there exists a map $T: \X \to \RR^{n_z}$ satisfying \eqref{eq:sylvester} with
\begin{subequations}
\label{eq_AB_gamma}
\begin{align}
 A &=\gamma \at = \gamma \diag\left(\at_1, \at_2, \ldots, \at_{n_y}\right) \in \RR^{n_z \times n_z}, \\ B& = \diag\left(\bt_1, \bt_2, \ldots, \bt_{n_y}\right) \in \RR^{n_z \times n_y},
\end{align}
\end{subequations}
that has the four properties below:
\begin{enumerate}[label=(P\arabic*)]
\item $T$ is solution to \eqref{eq:sylvester} on $\X$;
\item $T$ is Lipschitz injective on $\X$, i.e., there exists $c_I > 0$ (independent of $\gamma)$ such that for all $(x_a, x_b) \in \X \times \X$,
 \begin{equation}
 \label{eq:lips_inject}
      \|T(x_a) - T(x_b)\| \geq c_I \gamma^{\overline{m} -1} \|x_a - x_b\|,
 \end{equation}
 with {\footnotesize$c_I = c_N\left(c_c c_o- \displaystyle\max_{i=\overline{1,n_y}}\|\bt_i\| c_h c_f \frac{\gamma \displaystyle\max_{i=\overline{1,n_y}}((\|\at_i\|c_f)^{m_i})}{1 - \gamma\displaystyle\max_{i=\overline{1,n_y}}\|\at_i\|c_f} \right)$},
 where $c_N > 0$ is a constant depending on the norm and $c_c > 0$ is the bound of the inverse of the (constant) controllability matrix of $(\at_i,\bt_i)$ (see the Appendix);
    \item $T$ is Lipschitz on $\RR^{n_x}$, i.e., there exists $c_L > 0$ such that for all $(x_a, x_b) \in \RR^{n_x} \times \RR^{n_x}$,
 \begin{equation}
 \label{eq:lips}
      \|T(x_a) - T(x_b)\| \leq c_L \|x_a - x_b\|;
 \end{equation}
 \item There exists a map $T^*: \RR^{n_z} \to \RR^{n_x}$ such that
\begin{subequations}\label{eq:propTs}
    \begin{align}
    T^*(T(x))& = x, \quad \forall x \in \X,\\
  \|T^*(z_a) - T^*(z_b)\| &\leq \frac{c}{\gamma^{\overline{m}-1}}\|z_a - z_b\|,\notag \\&\quad \forall (z_a, z_b) \in \RR^{n_z} \times \RR^{n_z},
\end{align}
\end{subequations}
with some $c > 0$.
\end{enumerate}
\end{lemma}

Note that the scalars in Lemma \ref{lemma} are obtained from Assumption \ref{ass_sys} and the choice of $(A,B)$, and $\gamma$. They can be picked very conservatively. Also, if for all $x \in \X$ compact and for all $i \in \NN$, we have $(\underbrace{f^{-1} \p \ldots \p f^{-1}}_{i \text{ times}})(x) \in \X$, then $T$ can be shown to be unique following \cite[Theorem 4]{brivadisCDC}.

\begin{proof}
    First, the existence of $T: \X \to \RR^{n_z}$ for all $x \in \X$ follows from \cite[Theorem 2]{brivadisCDC} under Item \ref{ass_inv} of Assumption \ref{ass_sys}, given by the closed form
\begin{equation}\label{eq:T}
    T(x) = \sum_{i = 0}^{+\infty}A^i B (h \p \underbrace{f^{-1} \p \ldots \p f^{-1}}_{i+1 \text{ times}})(x), \quad \forall x \in \X.
\end{equation}
Second, the Lipschitz injectivity of $T$ on $\X$ can be proven for \eqref{eq:T} under Items \ref{ass_fh} and \ref{ass_obs} of Assumption \ref{ass_sys} by adapting \cite[Proof of Theorem 3]{baokkltac} for time-invariant systems with $\gamma \in (0,1]$ such that
\begin{multline*}
    0 < \gamma < \gamma^\star := \min\bigg\{\frac{1}{\|\at\|},\frac{1}{\displaystyle\max_{i=\overline{1,n_y}}\|\at_i\| c_f},\\ \frac{c_c c_o}{\displaystyle\max_{i=\overline{1,n_y}}\|\at_i\|c_fc_c c_o + \displaystyle\max_{i=\overline{1,n_y}}\|\bt_i\|c_h c_f\displaystyle\max_{i=\overline{1,n_y}}((\|\at_i\|c_f)^{m_i})}\bigg\}.
\end{multline*}
More details are given in the Appendix. Third, we prove the Lipschitzness of $T$ given by \eqref{eq:T} on $\X$. From Item \ref{ass_fh} of Assumption \ref{ass_sys}, it follows that since $\gamma \displaystyle\max_{i=\overline{1,n_y}}\|\at_i\|c_f<1$, for all $(x_a,x_b) \in \X \times \X$,
\begin{align*}
\|T(x_a)& - T(x_b)\|
\\
&\leq \sum_{j=0}^{+\infty}(\gamma \displaystyle\max_{i=\overline{1,n_y}}\|\at_i\|)^j \displaystyle\max_{i=\overline{1,n_y}}\|\bt_i\| c_h c_f^{j+1}\|x_a - x_b\|\\ & = \frac{\displaystyle\max_{i=\overline{1,n_y}}\|\bt_i\| c_h c_f}{1 - \gamma\displaystyle\max_{i=\overline{1,n_y}}\|\at_i\|c_f}\|x_a - x_b\|:=c_L\|x_a - x_b\|,
\end{align*}
which is fixed once we fix $\gamma$. Finally, the existence of $T^*$ satisfying \eqref{eq:propTs} is deduced from \eqref{eq:lips_inject} by applying \cite[Theorem 1]{baokkltac}, which is based on \cite{mcshane}.
\end{proof}

At the end of this part, we know to pick $\gamma$ sufficiently small so that $T$ is left-invertible and there exists $T^*$ with the said properties.

\subsection{Interval Observer Design}
To design our KKL-based interval observer, the following assumption on $(w_k)_{k \in \NN}$ is classical in interval observer design and is frequently satisfied in practice. 
\begin{assumption}\label{ass_dw} There exist known sequences $\left(w_k^+,w_k^-\right)_{k \in \NN}$ such that the noise $(w_k)_{k \in \NN}$ satisfies $w_k^- \leq w_k \leq w_k^+$ for all $k \in \NN$.
\end{assumption}

We propose for system \eqref{eq:sysz} in the $z$-coordinates the interval observer candidate
\begin{subequations}\label{eq:obsz}
\begin{equation}\label{eq:hatznonlin}
 \left\{ \begin{array}{@{}r@{\;}c@{\;}l@{}}
        \hat{z}_{k+1}^+ &=& R_{k+1}AR_k^{-1} \hat{z}_k^+ + R_{k+1}B y_k\\
        &&{} +\left(R_{k+1}B\right)^\ominus w_k^+ - \left(R_{k+1}B\right)^\oplus w_k^-\\
        \hat{z}_{k+1}^- &=& R_{k+1}AR_k^{-1} \hat{z}_k^- + R_{k+1}B y_k \\
        &&{}+ \left(R_{k+1}B\right)^\ominus w_k^- - \left(R_{k+1}B\right)^\oplus w_k^+,
    \end{array}\right.
\end{equation}
with the initial conditions
\begin{align}\label{eq:init}
        \hat{z}_{0}^+= R_{0}^\oplus z_0^+ - R_0^\ominus z_0^-, \quad
        \hat{z}_{0}^-= R_{0}^\oplus z_0^- - R_0^\ominus z_0^+,
\end{align}
in which, component-wise for all $i=\overline{1,n_z}$,
\begin{align}
z_{0,i}^+&= \min\left\{T\left(x_0^+\right)_{i},T\left(x_0^-\right)_i\right\}\notag\\&\hspace{3cm}{}+c_L\displaystyle\max_{j=\overline{1,n_x}}\left(x_{0,j}^+-x_{0,j}^-\right), 
 \label{eq:initznonlinp}\\
z_{0,i}^-&= \max\left\{T\left(x_0^+\right)_i,T\left(x_0^-\right)_i\right\}\notag\\&\hspace{3cm}{}-c_L\displaystyle\max_{j=\overline{1,n_x}}\left(x_{0,j}^+-x_{0,j}^-\right),\label{eq:initznonlinm}
\end{align}
and the bounds for $k \geq 1$
\begin{equation}\label{eq:initnonlin}
      z_{k}^+= S_{k}^\oplus\hat{z}_k^+ - S_k^\ominus\hat{z}_k^-, \quad
        z_{k}^-= S_{k}^\oplus\hat{z}_k^- - S_k^\ominus\hat{z}_k^+,
\end{equation}
where $(R_k)_{k \in \NN}$ such that for all $k \in \NN$, $R_{k+1}AR_k^{-1}$ is a non-negative Schur constant matrix and $S_k=R_k^{-1}$ following Lemma \ref{lem_Rk}. Then the estimate is recovered in the $x$-coordinates by
recovering the bounds at all times using, for all $i=\overline{1,n_x}$, 
\begin{align}
        x_{k,i}^+ &=\min\left\{T^*\left(z_k^+\right)_i,T^*\left(z_k^-\right)_i\right\}\notag\\&\hspace{2.5cm}{}+\frac{c}{\gamma^{\overline{m}-1}}\displaystyle\max_{j=\overline{1,n_z}}\left(z_{k,j}^+-z_{k,j}^-\right),\label{eq:xknonlinp} \\ 
    x_{k,i}^- &= \max\left\{T^*\left(z_k^+\right)_i,T^*\left(z_k^-\right)_i\right\}\notag\\&\hspace{2.5cm}{}-\frac{c}{\gamma^{\overline{m}-1}}\displaystyle\max_{j=\overline{1,n_z}}\left(z_{k,j}^+-z_{k,j}^-\right),\label{eq:xknonlinm}
\end{align}
\end{subequations}
with $m_i$ defined in Definition \ref{def_obs}. Note that due to the nonlinearity in system \eqref{eq:sysx}, it typically needs to be transformed into one of higher dimension, namely $n_z \geq n_x$, for the transformation $T$ to be left-invertible (in our case, $n_z$ is defined in Theorem \ref{theo}). Therefore, we cannot write the observer dynamics in the $x$-coordinates. 
\begin{remark}\label{rem}
First, for each $i=\overline{1,n_z}$, the term $\min\left\{T^*\left(z_k^+\right)_i,T^*\left(z_k^-\right)_i\right\}$ in \eqref{eq:xknonlinp} can be replaced with flexibility using either $T^*\left(z_k^+\right)_i$, $T^*\left(z_k^-\right)_i$, or $\max\left\{T^*\left(z_k^+\right)_i,T^*\left(z_k^-\right)_i\right\}$. Similarly, $\max\left\{T^*\left(z_k^+\right)_i,T^*\left(z_k^-\right)_i\right\}$ in \eqref{eq:xknonlinm} can be interchanged with $T^*\left(z_k^+\right)_i$, $T^*\left(z_k^-\right)_i$, or $\min\left\{T^*\left(z_k^+\right)_i,T^*\left(z_k^-\right)_i\right\}$. Second, the conservatism arising from the selection of $c_L$ in \eqref{eq:initznonlinp}-\eqref{eq:initznonlinm} is not a significant concern, as the impact of the initial conditions on the interval width will be forgotten over time. Last, to simplify exposition, the sequence of additive disturbance $(d_k)_{k \in \NN}$, with known bounds $(d_k^-)_{k \in \NN}$ and $(d_k^+)_{k \in \NN}$ such that $d_k^- \leq d_k \leq d_k^+$ for all $k \in \NN$, is not present in system \eqref{eq:sysx}. In the presence of such a $(d_k)_{k \in \NN}$, still with $T$ satisfying \eqref{eq:sylvester}, system \eqref{eq:sysz} becomes
    \begin{multline}
        z_{k+1} = A z_k + B y_k - Bw_k \\
        + T\left(f(x_k) + d_k\right) - T\left(f(x_k)\right),
    \end{multline}
   with $x_k$ solution to $x_{k+1} = f(x_k) + d_k$ and with $y_k = h(x_k) + w_k$.
    Thanks to the Lipschitzness of $T$ exhibited in \eqref{eq:lips}, we have for all $x \in \RR^{n_x}$ and for all $k \in \NN$, $\left\|T(f(x) + d_k) - T(f(x))\right\| \leq 
        c_L\max\left\{\left\|d_k^+\right\|,\left\|d_k^-\right\|\right\}$.
    Thus, for all $x \in \RR^{n_x}$ and for all $k \in \NN$,
    \begin{multline}
       -\displaystyle\max\left\{\max_{i=\overline{1,n_x}}\left(\left|d_{k,i}^+\right|\right),\max_{i=\overline{1,n_x}}\left(\left|d_{k,i}^-\right|\right)\right\}E_{n_z} 
       \\
       \leq T(f(x) + d_k) - T(f(x))\leq
        \\
     \displaystyle\max\left\{\max_{i=\overline{1,n_x}}\left(\left|d_{k,i}^+\right|\right),\max_{i=\overline{1,n_x}}\left(\left|d_{k,i}^-\right|\right)\right\}E_{n_z}. 
    \end{multline}
    Consequently, the result of this section can be extended straightforwardly to the case where $(d_k)_{k \in \NN}$ is present.
\end{remark}

In the favorable case where the inverse map $T^*$ happens to be mixed monotone as in \cite[Definition 4]{Yang-19}, recovering the bounds at all times in the $x$-coordinates is obvious \cite[Lemma 2]{Dinhetal:2023:IFACWC}, thus \eqref{eq:xknonlinp} and \eqref{eq:xknonlinm} can be simplified to
\begin{equation}\label{eq:mixmonot}
    x_k^+ =T_d^*\left(z_k^+,z_k^-\right), \quad 
    x_k^- = T_d^*\left(z_k^-,z_k^+\right),
    \end{equation}
    where $T_d^*$ is a decomposition function of $T^*$, which is less conservative than 
    \eqref{eq:obsz}. However, such a design is only situational since $T^*$ is not guaranteed to be mixed monotone, and our design \eqref{eq:obsz} does not require this property from $T^*$.
\begin{theorem}\label{theo}
Let system \eqref{eq:sysx} satisfy Assumptions \ref{ass_sys} and \ref{ass_dw}. Define $n_z = \sum_{i = 1}^{n_y} m_i$ with $m_i$ in Definition \ref{def_obs}. Consider for each $i \in \left\{1, 2, \ldots, n_y\right\}$ a controllable pair $\left(\at_i, \bt_i\right) \in \mathbb{R}^{m_i\times m_i} \times \mathbb{R}^{m_i}$ where $\at_i$ is a Schur matrix. Then, there exists a sequence of invertible real matrices $(R_k)_{k \in \NN}$ such that for all $k \in \NN$, $R_{k+1}AR_k^{-1}$ is a non-negative Schur constant matrix. Then, the dynamic extensions \eqref{eq:obsz} are an interval observer for system \eqref{eq:sysx}, with an arbitrarily fast convergence rate in the absence of $(w_k)_{k \in \NN}$.
\end{theorem}
\begin{proof}
First, apply Lemma \ref{lemma} to show the existence of $T$ and $T^*$ satisfying the conditions therein, so that the dynamic extensions \eqref{eq:obsz} are properly defined.
Because $x_0^- \leq x_0 \leq x_0^+$ and thanks to the Lipschitz property \eqref{eq:lips}, we have $\left\|T\left(x_0^+\right)-T\left(x_0\right)\right\| \leq c_L\left\|x_0^+-x_0\right\| \leq c_L\left\|x_0^+-x_0^-\right\|$ and $\left\|T\left(x_0\right)-T\left(x_0^-\right)\right\| \leq c_L\left\|x_0-x_0^-\right\| \leq c_L\left\|x_0^+-x_0^-\right\|$. So,
$\max_{i=\overline{1,n_z}}\left(\left|T\left(x_0^+\right)_i - z_{0,i}\right|\right) \leq c_L\max_{i=\overline{1,n_x}}\left(x^+_{0,i} - x^-_{0,i}\right)$, $\max_{i=\overline{1,n_z}}\left(\left|z_{0,i} - T\left(x_0^-\right)_i\right|\right) \leq c_L\max_{i=\overline{1,n_x}}\left(x^+_{0,i} - x^-_{0,i}\right)$. Consequently, component-wise for all $i=\overline{1,n_z}$, we have $-c_L\max_{j=\overline{1,n_x}}\left(x^+_{0,j} - x^-_{0,j}\right) 
    \leq T\left(x_0^+\right)_i - z_{0,i} 
    \leq c_L\max_{j=\overline{1,n_x}}\left(x^+_{0,j} - x^-_{0,j}\right)$,
and similarly $-c_L\max_{j=\overline{1,n_x}}\left(x^+_{0,j} - x^-_{0,j}\right) 
    \leq z_{0,i} - T\left(x_0^-\right)_i 
    \leq c_L\max_{j=\overline{1,n_x}}\left(x^+_{0,j} - x^-_{0,j}\right)$. Hence, component-wise for all $i=\overline{1,n_z}$,
\begin{multline*}
    T\left(x_0^+\right)_i-c_L\max_{j=\overline{1,n_x}}\left(x^+_{0,j} - x^-_{0,j}\right) \leq  z_{0,i}
    \\
\leq T\left(x_0^+\right)_i+ c_L\max_{j=\overline{1,n_x}}\left(x^+_{0,j} - x^-_{0,j}\right),
\end{multline*}
and correspondingly
\begin{multline*}
    T\left(x_0^-\right)_i-c_L\max_{j=\overline{1,n_x}}\left(x^+_{0,j} - x^-_{0,j}\right) \leq  z_{0_i}
\\\leq T\left(x_0^-\right)_i+ c_L\max_{j=\overline{1,n_x}}\left(x^+_{0,j} - x^-_{0,j}\right).
\end{multline*}
Therefore, defining $z_0^+$ and $z_0^-$ as in \eqref{eq:initznonlinp} and \eqref{eq:initznonlinm} implies that $z_0^- \leq z_0 \leq z_0^+$. From Lemma \ref{lem_pm}, it follows that $R_0^\oplus z_0^- - R_0^\ominus z_0^+ \leq R_0z_0 \leq R_0^\oplus z_0^+-R_0^\ominus z_0^-$. From \eqref{eq:initnonlin}, we get
    \begin{align}\label{eq:R0z0}
        \hat{z}_0^- \leq R_0z_0 \leq \hat{z}_0^+.
    \end{align}
    Next, consider the solutions $\left(z_k,\hat{z}_k^+,\hat{z}_k^-\right)_{k \in \NN}$ to the system
    \begin{equation}\label{eq:extensions}
\left\{\begin{array}{@{}r@{\;}c@{\;}l@{}}
        z_{k+1} &=& A z_k + B y_k - Bw_k\\
        \hat{z}_{k+1}^+ &=& R_{k+1}A R_k^{-1} \hat{z}_k^+ + R_{k+1}B y_k  \\&&{} + \left(R_{k+1}B\right)^\ominus w_k^+ - \left(R_{k+1}B\right)^\oplus w_k^-\\
        \hat{z}_{k+1}^- &=& R_{k+1}A R_k^{-1} \hat{z}_k^- + R_{k+1}B y_k \\&&{} + \left(R_{k+1}B\right)^\ominus w_k^- - \left(R_{k+1}B\right)^\oplus w_k^+.
    \end{array}\right.
\end{equation}
    Then, $R_{k+1}z_{k+1} = R_{k+1}AR_k^{-1}R_kz_k + R_{k+1}By_k
        -R_{k+1}Bw_k$.
    Thus, it follows that
    \begin{align*}
        \hat{z}_{k+1}^+&-R_{k+1}z_{k+1} = R_{k+1}AR_k^{-1}\left(\hat{z}_k^+-R_kz_k\right) \nonumber
        \\
        &{}\underbrace{+\left(R_{k+1}B\right)^\ominus w_k^+ - \left(R_{k+1}B\right)^\oplus w_k^-+R_{k+1}Bw_k}_{=p_k},\\
        R_{k+1}&z_{k+1}-\hat{z}_{k+1}^- = R_{k+1}AR_k^{-1}\left(R_kz_k-\hat{z}_k^-\right) \nonumber
        \\
        &{} \underbrace{- \left(R_{k+1}B\right)^\ominus w_k^- + \left(R_{k+1}B\right)^\oplus w_k^+-R_{k+1}Bw_k}_{=q_k}.
    \end{align*}
    From Lemma \ref{lem_pm}, we have $\left(R_{k+1}B\right)^\ominus w_k^+ -\left(R_{k+1}B\right)^\oplus w_k^- \geq -R_{k+1}Bw_k 
    \geq \left(R_{k+1}B\right)^\ominus w_k^--\left(R_{k+1}B\right)^\oplus w_k^+$.
    Hence, $p_k =\left(R_{k+1}B\right)^\ominus w_k^+ - \left(R_{k+1}B\right)^\oplus w_k^-+R_{k+1}Bw_k \geq 0$ and $q_k= - \left(R_{k+1}B\right)^\ominus w_k^- + \left(R_{k+1}B\right)^\oplus w_k^+-R_{k+1}Bw_k \geq 0$.
    Because the matrix $R_{k+1}AR_k^{-1}$ is non-negative, $p_k \geq 0$ and $q_k \geq 0$ for all $k \in \NN$, and $0 \leq \hat{z}_0^+-R_0z_0$ and $0 \leq R_0z_0 -\hat{z}_0^-$ (according to \eqref{eq:R0z0}), we can deduce that $\hat{z}_k^- \leq R_kz_k \leq \hat{z}_k^+$ for all $k \in \NN$. From \eqref{eq:initnonlin} and Lemma \ref{lem_pm}, it follows that 
\begin{align}
    z^-_k \leq z_k \leq z^+_k, \qquad \forall k \in \NN.
\end{align}
Because $x_k \in \X$ for all $k \in \NN$ and thanks to \eqref{eq:propTs}, we have
\begin{align*} \nonumber
    \left\|T^*\left(z_k^+\right)-x_k\right\| & = \left\|T^*\left(z_k^+\right)-T^*\left(T(x_k)\right)\right\|\\
    &\leq \frac{c}{\gamma^{\overline{m}-1}} \left\|z_k^+-T(x_k)\right\|\nonumber
    \\ \nonumber
    &{}=\frac{c}{\gamma^{\overline{m}-1}} \left\|z_k^+-z_k\right\|
\leq \frac{c}{\gamma^{\overline{m}-1}} \left\|z_k^+-z_k^-\right\|,
\end{align*}
and analogously $\left\|x_k - T^*\left(z_k^-\right)\right\|\leq \frac{c}{\gamma^{\overline{m}-1}} \left\|z_k^+-z_k^-\right\|$.
With the same arguments as above, we obtain for all $i=\overline{1,n_x}$, $    T^*\left(z_k^+\right)_i-\frac{c}{\gamma^{\overline{m}-1}}\displaystyle\max_{j=\overline{1,n_z}}\left(z_{k,j}^+-z_{k,j}^-\right)
     \leq x_{k,i}
   \leq T^*\left(z_k^+\right)_i+\frac{c}{\gamma^{\overline{m}-1}}\displaystyle\max_{j=\overline{1,n_z}}\left(z_{k,j}^+-z_{k,j}^-\right)$,
and $T^*\left(z_k^-\right)_i-\frac{c}{\gamma^{\overline{m}-1}}\displaystyle\max_{j=\overline{1,n_z}}\left(z_{k,j}^+-z_{k,j}^-\right)
     \leq x_{k,i} 
   \leq
    T^*\left(z_k^-\right)_i+\frac{c}{\gamma^{\overline{m}-1}}\displaystyle\max_{j=\overline{1,n_z}}\left(z_{k,j}^+-z_{k,j}^-\right)$.
From \eqref{eq:xknonlinp} and \eqref{eq:xknonlinm}, it follows that $x_k^- \leq x_k \leq x_k^+$ for all $k \in \NN$.
Finally, we deduce from \eqref{eq:hatznonlin} that, in the absence of $(w_k)_{k \in \NN}$,
\begin{equation}\label{eq:stable2}
        \hat{z}_{k+1}^+-\hat{z}_{k+1}^-=R_{k+1}AR_k^{-1}\left(\hat{z}_k^+-\hat{z}_k^-\right).
    \end{equation}
Note that $R_{k+1}AR_k^{-1}$ is Schur. Then from \eqref{eq:obsz} and the exponential
stability in the $z$-coordinates in \eqref{eq:stable2}, we have 
{\small\begin{align*}\nonumber
    \left\|x_k^+-x_k^-\right\| &\leq \frac{2c}{\gamma^{\overline{m}-1}} \left\|z_k^+-z_k^-\right\| + \left\|T^*\left(z_k^+\right)-T^*\left(z_k^-\right)\right\|
   \nonumber\\
   &\leq \frac{3c}{\gamma^{\displaystyle\max_{i=\overline{1,n_y}} m_i-1}}\|z_k^+-z_k^-\|\nonumber\\\nonumber
   & \leq \frac{3c}{\gamma^{\overline{m}-1}} c_1\left\|\hat{z}_k^+-\hat{z}_k^-\right\|
    \leq \frac{3c}{\gamma^{\overline{m}-1}} c_1c_2^k\left\|x_0^+-x_0^-\right\|,
\end{align*}}for some $c_1 > 0$ and $c_2 \in (0, 1)$. Besides, because $A$ given in \eqref{eq_AB_gamma} can be pushed arbitrarily close to $0$ by pushing $\gamma$ smaller, this gives us an interval observer with arbitrarily fast convergence as soon as allowed by Item \ref{ass_obs} of Assumption \ref{ass_sys} following \cite{baokkltac} (in the absence of $(w_k)_{k \in \NN}$).
\end{proof}

\section{An Illustrative Example}\label{sec_eg}
\begin{figure}
        \centering
\includegraphics[width=0.5\textwidth,height=0.25\textwidth]{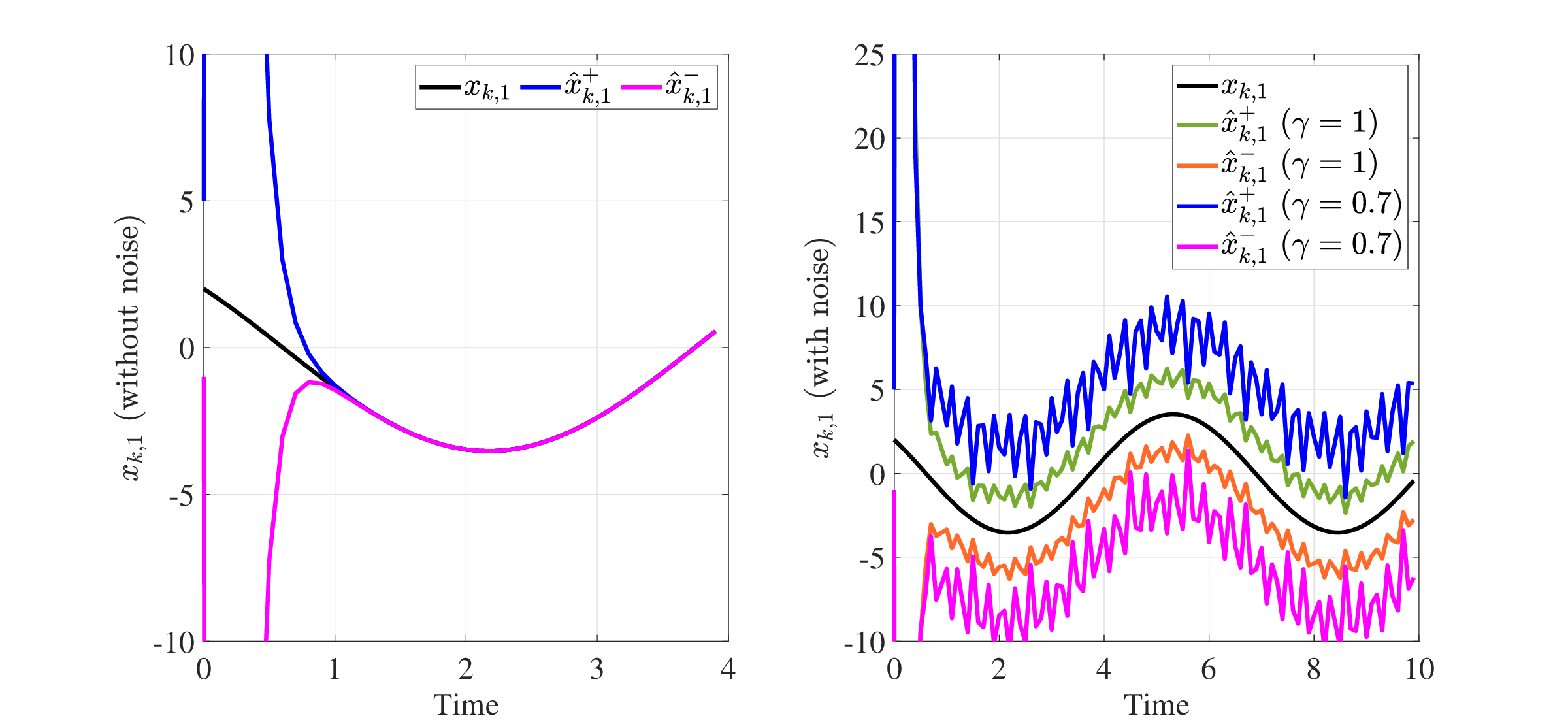}
        \caption{Results for $x_{k,1}$. Left: Convergence in the absence of noise; Right: Comparison between different observer parameters in the presence of noise.}       \label{fig:0noise}
    \end{figure}

Consider the system with dynamics and output:
\begin{subequations}\label{eq:syseg}
    \begin{align}
    \begin{pmatrix}
        x_{k+1,1}\\x_{k+1,2}
    \end{pmatrix}&=\begin{pmatrix}
        x_{k,1} - \tau x_{k,2}\\(1-\tau^2)x_{k,2} + \tau x_{k,1}
    \end{pmatrix},
     \\ y_k &= x_{k,1}^2 - x_{k,2}^2 + x_{k,1} + x_{k,2} + w_k.
\end{align}
\end{subequations}
First, this system results from the semi-implicit Euler discretization with rate $\tau > 0$ of the continuous-time system\begin{equation}\label{eq:sysegc}
    \begin{pmatrix}
        \dot{x}_1\\\dot{x}_2
    \end{pmatrix}=\begin{pmatrix}
        -x_2\\x_1
    \end{pmatrix}, \quad
      y = x_1^2 - x_2^2 + x_1 + x_2 + w.
\end{equation}
System \eqref{eq:sysegc} is known to be instantaneously observable of order $4$, i.e., the map $x \mapsto (y,\dot{y},\ddot{y},\dddot{y})$ is injective in $x$ \cite{brivadisCDC}. We then conjecture that when $\tau$ is small, the discrete-time equivalence with the same order $\overline{m} = 4$ holds for the discrete-time system \eqref{eq:syseg}. Based on the linear dynamics and the quadratic output, picking $\tilde{A} = \diag(\lambda_1,\lambda_2,\lambda_3,\lambda_4)$ and $B = E_4$ (controllable), we look for $T$ of the form
\begin{subequations}
    \begin{equation}
            T(x) = (T_{\lambda_1}(x), T_{\lambda_2}(x), T_{\lambda_3}(x), T_{\lambda_4}(x)),
        \end{equation}
    where each line has the form
        \begin{equation}
            T_{\lambda_i}(x) = a_{\lambda_i}x_1^2 + b_{\lambda_i}x_2^2 + c_{\lambda_i}x_1x_2 + d_{\lambda_i}x_1 + e_{\lambda_i}x_2,
        \end{equation}
\end{subequations}
    where each parameter set depends on $\lambda_i$ following a relation obtained by solving \eqref{eq:sylvester} (see \cite{brivadisCDC} for more details with a similar example). We pick $\lambda_1 = 0.1$, $\lambda_2 = 0.2$, $\lambda_3 = 0.3$, and $\lambda_4 = 0.4$ and take $\tau = 0.1$ (s). Let us assume some bounded set $\X$ where solutions of interest remain (even in backward time) and the constants could be taken on this set. More particularly, $c_L$ can be approximately taken as the upper bound of the derivative of $T$ with respect to $x$, or the norm of the Jacobian matrix of $T$, for $x \in \X$, and $c$ can be taken as $\frac{1}{c_I}$ where all constants therein are taken on $\X$. It is seen that $T$ is Lipschitz injective with $\gamma \leq 1$, and we invert it numerically thanks to a nonlinear optimization-based solver. 
    In the absence of $(w_k)_{k \in \NN}$, as in Figure \ref{fig:0noise}-left, the interval observer behaves like a pair of KKL observers, with peaking followed by (arbitrarily fast) exponential convergence.
In the case of noise $w_k=0.2\cos(20k)$, we choose $w_k^+=\max\left\{0.2\cos(20k),\frac{0.5}{k^2}\right\}$, $w_k^-=\min\left\{0.2\cos(20k),\frac{0.5}{k^2}\right\}$, and compare between the choices of $\gamma = 1$ and $\gamma = 0.7$. Results suggest that a smaller $\gamma$, which is linked to a higher degree of nonlinearity in the system, leads to higher conservativeness in the bounds---see Figure \ref{fig:0noise}-right.

\section{Conclusion}\label{sec_conclu}
We propose a theoretical interval observer design based on the KKL framework for nonlinear discrete-time systems, without any assumption on the structure of the dynamics and output. It is also expected that similar results can be obtained in continuous time, based on \cite{brivadisRemarks,bernardNonautonomousKKL}. Future developments include improving estimation performance by adapting the observer parameters to the operating condition of the system.
\appendix
The proof of (P2) in Lemma \ref{lemma} is based on the form of $T$ in \eqref{eq:T}. First, pick $\gamma\in (0,1]$ small enough to have $\gamma\|\at\|<1$. Consider a solution $T$ of \eqref{eq:sylvester} for $(A,B)$ given in \eqref{eq_AB_gamma}. Then, 
$$
T(x) = (T_1(x), T_2(x), \ldots, T_i(x), \ldots, T_{n_y}(x)),
$$ 
where for each $i\in \{1,2,\ldots,n_y\}$, $T_i$ is solution to \eqref{eq:sylvester} with $(A,B)$ replaced by $(\gamma \tilde{A}_i,\tilde{B}_i)$. For each $i \in \{1,2,\ldots,n_y\}$, for all $(x_a, x_b) \in \X \times \X$, $T_i(x_a) - T_i(x_b)$ is written as
\begin{equation*}
    T_i(x_a) - T_i(x_b) = \left(\T_i(x_a) - \T_i(x_b)\right) + \left(\R_i(x_a) - \R_i(x_b)\right),
\end{equation*}
where
{\small\begin{equation*}
\begin{array}{@{}r@{\;}c@{\;}l@{}}
\T_i(x_a) - \T_i(x_b)& =& \displaystyle\sum_{j=0}^{m_i-1}(\gamma \at_i)^j\bt_i((h_i \p \underbrace{f^{-1} \p \ldots \p f^{-1}}_{j+1 \text{ times}})(x_a)\\&& {}- (h_i \p \underbrace{f^{-1} \p \ldots \p f^{-1}}_{j+1 \text{ times}})(x_b)) \\&=& \D_i(\gamma)\C_i (\OO_i(x_a) - \OO_i(x_b)),\\
\R_i(x_a) - \R_i(x_b)& =& \displaystyle\sum_{j=m_i}^{+\infty}(\gamma \at_i)^j\bt_i((h_i \p \underbrace{f^{-1} \p \ldots \p f^{-1}}_{j+1 \text{ times}})(x_a)\\&&{} - (h_i \p \underbrace{f^{-1} \p \ldots \p f^{-1}}_{j+1 \text{ times}})(x_b)),
\end{array}
\end{equation*}}
where $\D_i(\gamma) = \diag(1, \gamma, \gamma^2, \ldots, \gamma^{m_i-1})$ and $\C_i = \begin{pmatrix}
    \bt_i & \at_i \bt_i & \at_i^2 \bt_i & \ldots & \at_i^{m_i-1}\bt_i
\end{pmatrix}$ is the controllability matrix of the pair $(\tilde{A}_i,\tilde{B}_i)$.
Then, for $\gamma$ such that $\gamma\max_i\|\at_i\|c_f < 1$, exploiting Item \ref{ass_fh} of Assumption \ref{ass_sys}, for all $i \in \{1,2,\ldots,n_y\}$ and for all $(x_a, x_b) \in \X \times \X$,
\begin{equation*}
        \|\R_i(x_a) - \R_i(x_b)\| \leq \|\bt_i\| c_h c_f \frac{(\gamma\|\at_i\|c_f)^{m_i}}{1 - \gamma\|\at_i\|c_f}\|x_a - x_b\|.
\end{equation*}
Since the pairs $(\at_i, \bt_i)\in \RR^{m_i\times m_i}\times \RR^{m_i}$ are controllable, there exists $c_c > 0$ such that $\|\C_i^{-1}\| \leq \frac{1}{c_c}$ for all $i \in \{1, 2, \ldots, n_y\}$. Next, we deduce that for all $i \in \{1,2,\ldots,n_y\}$ and for all $(x_a, x_b) \in \X \times \X$,
\begin{equation*}
    \begin{split}
        \|\T_i(x_a) - \T_i(x_b)\| \geq \gamma^{m_i-1} c_c \|\OO_i(x_a) -  \OO_i(x_b)\|.
    \end{split}
\end{equation*}
So, for all $i \in \{1,2,\ldots,n_y\}$ and for all $(x_a, x_b) \in \X \times \X$,
{\small\begin{align*}
        \|T_i(x_a) - T_i(x_b)\| 
        &\geq \gamma^{m_i-1} \bigg(c_c \|\OO_i(x_a) -  \OO_i(x_b)\|\\& \quad- \|\bt_i\| c_h c_f \frac{\gamma(\|\at_i\|c_f)^{m_i}}{1 - \gamma\|\at_i\|c_f}\|x_a - x_b\|\bigg).
\end{align*}}
Now, since $\gamma \in (0,1]$ and thanks to Item \ref{ass_obs} of Assumption \ref{ass_sys}, if we concatenate the outputs, there exists a constant $c_N > 0$ such that for all $(x_a, x_b) \in \X \times \X$, we have 
\begin{multline*}
        \|T(x_a) - T(x_b)\|\geq  c_N \gamma^{\max_i m_i-1}  \\\times\bigg(c_c c_o- \max_i\|\bt_i\| c_h c_f \frac{\gamma \max_i((\|\at_i\|c_f)^{m_i})}{1 - \gamma\max_i\|\at_i\|c_f} \bigg) \|x_a - x_b\|.
\end{multline*}
If we pick $\gamma \in (0,1]$ as in the proof of Lemma \ref{lemma}, there exists $c > 0$ (independent of $\gamma$) such that for all $(x_a, x_b) \in \X \times \X$, we have \eqref{eq:lips_inject}.


\bibliographystyle{IEEEtran}
\bibliography{ref}
\end{document}